\documentclass[12pt,a4paper]{article}
\usepackage{amsmath,amssymb}
\title{Nonperturbative deformation of D-brane states by the world sheet noncommutativity}
\author{Manabu Irisawa}
\date{February, 2008}
\def\_{ \hspace{.25ex} }
\def\nn{ \nonumber }
\def\Z{ \mathbb{Z} }

\def\R{ \mathbb{R} }
\def\d{ \mathrm{d} }
\def\pd{ \partial }
\def\vac{ | 0 \rangle }
\def\cvac{ \langle 0 | }
\def\NormalOrdering{ \text{\large\lower0.15ex\hbox{\makebox[.4em]{\rm :}}} }
\def\X{ X }
\def\Xp{ \X_{+} }
\def\Xm{ \X_{-} }
\def\tX{ \widetilde{\X} }
\def\tXp{ \tX_{+} }
\def\tXm{ \tX_{-} }
\def\Step{ \alpha }
\def\tStep{ \tilde{\Step} }
\def\Cx{ x_{0} }
\def\Cm{ p_{0} }

\def\Aux{ \xi }
\def\tAux{ \tilde{\Aux} }
\def\RTheta#1{ \vartheta_{#1} }
\def\NCParam{ \lambda }
\def\NCProd{ \ast }
\def\nlNCProd{ \ast }
\def\NCBS#1{ | #1 \rangle_{\NCProd} }
\def\cNCBS#1{ {\!}_{\NCProd}\langle #1 | }
\def\NCBracket#1#2{ {\!}_{\NCProd}\langle #1 | #2 \rangle_{\NCProd} }
\def\Vertex#1{ V^{\NCProd}_{#1} }
\makeatletter
\def\metadata#1{ \begin{flushright} \@date \\ arXiv:#1 \end{flushright} }
\def\thetitle{ \begin{center} \Large \bfseries{\@title} \end{center} }
\def\theauthor{ \begin{center} \large \@author \end{center} }
\newenvironment{address}{ \begin{center} \setlength{\baselineskip}{1.5em} }{ \end{center} }
\makeatother
\begin{document}
%
%#################################################
\metadata{0802.3292}
\thetitle
\vspace{.5em}
\theauthor
\begin{address}
{\it
Department of Physics,
Tokyo Metropolitan University, \\
1-1 Minami-osawa, Hachioji-shi,
Tokyo 192--0397, JAPAN \\
}
{\small \tt irisawa@kiso.phys.metro-u.ac.jp}
\end{address}
\vspace{.5em}
\begin{abstract}
By introducing the noncommutativity in the world sheet,
we discuss a modification of the D-brane states in the closed string theory.
In particular we show
how the world sheet noncommutativity induces a nonperturbative effect to the D-brane states.
\end{abstract}
%
%#################################################
\section{Introduction}
The noncommutative spacetime provided by the string theory \cite{Seiberg_Witten}
has been discussed as a possible basis of the theory of quantum gravity \cite{NC_Field_Theory}.
In recent years, it has been shown that
the noncommutative spaces can be constructed by the help of the concept of Hopf algebra
\cite{Hopf_Algebra_Procedure}
from certain commutative spaces which have Lie algebraic symmetries.
We can consider the string theory, the D-brane dynamics and the M-theory on the noncommutative space,
in which the noncommutativity plays an important role
in the description of their nonperturbative aspects.
\par
From this point of view, it will be interesting to investigate
the effect of the noncommutativity of the string world sheet.
In fact the two dimensional noncommutative conformal field theories
have been discussed in various aspects,
such as the perturbative approach \cite{Perturbative_NC_World_Sheet},
the harmonic oscillator formalism \cite{Hatzinikitas_Smyrnakis} and
the derivation of the star product by virtue of the Hopf algebra \cite{Lizzi_Vaidya_Vitale}.
In spite of many efforts, however, we are still far from obtaining some interesting results. 
\par
One of the main difficulties of these approaches owes to the fact that
the noncommutativity of the world sheet introduces a dimensional constant
which breaks explicitly the conformal symmetry of the theory.
To recover the theory from this problem,
one must commit oneself to many complicated formulae expanded into perturbative series.
\par
On the other hand the world sheet noncommutativity possesses a notable feature, {\it i.e.},
it does not affect the product of the left-movers, which are holomorphic,
nor the product of the right-movers, which are anti-holomorphic, of the strings.
Therefore the influence of the world sheet noncommutativity appears only in the product
between the left-movers and the right-movers in the closed string theory.
This means, for example, that
the world sheet noncommutativity does not change the theory of open strings at all.
Hence the world sheet noncommutativity is nothing to do with the spacetime quantization.
In other words we must quantize the spacetime independent from the world sheet noncommutativity.
\vspace{\baselineskip}
\par
Considering the above background into account, we would like to study, in this note,
some nonperturbative effects of the world sheet noncommutativity to the D-brane states.
To this end
we assume that all physical quantities can be represented by the string coordinates \( \X^{\mu}(\sigma,\tau) \)
and that the effect of the world sheet noncommutativity appears only through their products.
Moreover we assume that,
although the left-moving string coordinate \( \X^{\mu}(\sigma+i\tau) \)
and the right-moving string coordinate \( \tX^{\mu}(\sigma-i\tau) \) do not commute any more,
their components \( \Step^{\mu}_{n} \) and \( \tStep^{\mu}_{n} \) commute, so that
the world sheet noncommutativity does not change the conventional quantization rule of the spacetime.
We will see that these assumptions enable us to put forward our investigation nonperturbatively.
At the same time
we must emphasize that we leave aside the problem of the breaking of the conformal symmetry.
\par
In order to see the effect of the world sheet noncommutativity,
the D-brane is a convenient object to study because the D-brane boundary state
consists of a product of the left and right moving string coordinates.
The star product associated with the world sheet noncommutativity is introduced locally.
Nevertheless we will see that it produces a nonlocal deformation of the D-brane boundary state
as a nonperturbative effect.
In the D-brane correlators, for example, the noncommutativity parameter \( \NCParam \) appears
through the elliptic functions as their modular parameter.
All such results suggest that the world sheet noncommutativity
changes the topology of the world sheet itself.
\par
Since the D-brane is a much-discussed object and many results are accumulated,
we should be able to compare our results with them.
Although we do not even consider the fermionic partners and the ghosts in this note,
our results are enough to see the effect of the world sheet noncommutativity.
\vspace{\baselineskip}
\par
The organization of this note is as follows.
In section 2, we define the star product and consider the noncommutativity in the world sheet.
In section 3, we evaluate correlators of
the D-brane states as deformed by the world sheet noncommutativity.
Finally, we summarize the results in section 4.
%
%
%#################################################
\section{World sheet noncommutativity}
In two dimensional space,
the \( \NCProd \)-product 
can be represented by the differential operators of the coordinate as follows;
\begin{equation}
\label{EQN:Local_Star_Product}
	f(\sigma,\tau) \NCProd g(\sigma,\tau)
	=	\lim_{(\sigma',\tau') \to (\sigma,\tau)}
		e^{- \NCParam ( \pd_{\sigma} \pd_{\tau'} - \pd_{\tau} \pd_{\sigma'} )}
		f(\sigma,\tau) \, g(\sigma',\tau')
\, ,
\end{equation}
with only one noncommutativity parameter \( \NCParam \).
The commutator of the coordinates with respect to this product gives a constant value
which means noncommutative space:
\begin{equation*}
	\left[ \, \tau, \sigma \, \right]_{\NCProd}
	=	2 \NCParam
\, .
\end{equation*}
In order to investigate the physical aspects, however,
we have to discuss the theory of fields at different points
and so we give a versatility to the \( \NCProd \)-product which acts on multiple different points.
In the case of only two points,
we extend it by the copy of (\ref{EQN:Local_Star_Product}) without taking the limitation
\( (\sigma',\tau') \to (\sigma,\tau) \):
\begin{equation}
\label{EQN:Nonlocal_Star_Product}
	f(\sigma,\tau) \nlNCProd g(\sigma',\tau')
	=	e^{- \NCParam ( \pd_{\sigma} \pd_{\tau'} - \pd_{\tau} \pd_{\sigma'} )}
		f(\sigma,\tau) \, g(\sigma',\tau')
\, .
\end{equation}
Since this \( \nlNCProd \)-product is compatible with the trigonometric functions
\begin{equation*}
% \label{EQN:Product_of_Trigonometric_Functions}
	e^{i m \sigma + n \tau} \nlNCProd e^{i m' \sigma' + n' \tau'}
	=	e^{- i \NCParam ( mn' - nm' )} \,
		e^{i m \sigma + n \tau} e^{i m' \sigma' + n' \tau'}
\qquad \forall \,
	m, n, m'\!, n' \in \Z
\, ,
\end{equation*}
it is possible to define any product of fields at different points through their Fourier expansion.
In fact we can convince ourselves that the associativity
\begin{align*}
% \label{EQN:Associativity}
&	e^{i m_1 \sigma_1 + n_1 \tau_1} \nlNCProd
	\left( \,
		e^{i m_2 \sigma_2 + n_2 \tau_2} \nlNCProd
		e^{i m_3 \sigma_3 + n_3 \tau_3}
	\, \right)
\nn \\
&	=	e^{- i \NCParam ( m_2 n_3 - n_2 m_3 )} \,
		e^{i m_1 \sigma_1 + n_1 \tau_1} \nlNCProd
		\left( \,
			e^{i m_2 \sigma_2 + n_2 \tau_2}
			e^{i m_3 \sigma_3 + n_3 \tau_3}
		\, \right)
\nn \\
&	=	e^{- i \NCParam ( m_1 n_2 - n_1 m_2 )} \,
		e^{- i \NCParam ( m_2 n_3 - n_2 m_3 )} \,
		e^{- i \NCParam ( m_1 n_3 - n_1 m_3 )} \,
		e^{i m_1 \sigma_1 + n_1 \tau_1} \,
		e^{i m_2 \sigma_2 + n_2 \tau_2} \,
		e^{i m_3 \sigma_3 + n_3 \tau_3}
\nn \\
&	=	\left( \,
			e^{i m_1 \sigma_1 + n_1 \tau_1} \nlNCProd
			e^{i m_2 \sigma_2 + n_2 \tau_2}
		\, \right)
		\nlNCProd e^{i m_3 \sigma_3 + n_3 \tau_3}
\, 
\end{align*}
is preserved, which is the sufficient property for our purpose.
\vspace{\baselineskip}
\par
Now we focus our attention to the string theory. It is described by the string coordinate,
which is either holomorphic or anti-holomorphic function
of the world sheet coordinate \( (\sigma,\tau) \).
Due to this fact an evaluation of the contribution of the world sheet noncommutativity is not difficult.
\par
Let \( f_a(\sigma+i\tau) \) and \( g_a(\sigma-i\tau) \) be a holomorphic and an anti-holomorphic functions
which can be expanded into the series of the complex variable \( z = e^{i(\sigma+i\tau)} \)
and its conjugate, respectively, as follows
\begin{equation*}
	f_a(\sigma+i\tau)
	=	\sum_{n \in \Z} f_{a,n} \, e^{in(\sigma+i\tau)}
\ , \quad
	g_a(\sigma-i\tau)
	=	\sum_{n \in \Z} g_{a,n} \, e^{in(\sigma-i\tau)}
\ .
\end{equation*}
Then the \( \nlNCProd \)-product (\ref{EQN:Nonlocal_Star_Product}) gives the following formulae,
\begin{align}
\label{EQN:Holomorphic_vs_Holomorphic}
	f_a(\sigma+i\tau) \nlNCProd f_b(\sigma'+i\tau')
&	=	f_a(\sigma+i\tau) \, f_b(\sigma'+i\tau')
\, ,
\\[.6em]
\label{EQN:Antiholomorphic_vs_Antiholomorphic}
	g_a(\sigma-i\tau) \nlNCProd g_b(\sigma'-i\tau')
&	=	g_a(\sigma-i\tau) \, g_b(\sigma'-i\tau')
\, ,
\\
\label{EQN:Holomorphic_vs_Antiholomorphic}
	f_a(\sigma+i\tau) \nlNCProd g_b(\sigma'-i\tau')
&	=	\int_{0}^{2\pi} \!\!\! \int_{0}^{2\pi}
		\frac{\d \sigma_1 \d \sigma_2}{4 \pi \NCParam} \,
		e^{i(\sigma_1-\sigma)(\sigma_2-\sigma')/2\NCParam} \,
		f_a(\sigma_1+i\tau) \, g_b(\sigma_2-i\tau')
\, .
\end{align}
In the same point limit this formalism has been already mentioned in \cite{Weinstein}. 
We learn, from these results,
that any product of two holomorphic functions, as well as two anti-holomorphic functions,
is not affected by the world sheet noncommutativity.
In other words the world sheet noncommutativity does not violate the analytic properties of each field.
We have to attend our concern to the product only between a holomorphic and an anti-holomorphic components.
\vspace{\baselineskip}
\par
On the basis of definitions above,
we want to study the free closed string theory on the noncommutative world sheet.
The spacetime coordinates of a closed string are defined by
\begin{gather*}
\label{EQN:Fundamental_Objects}
	\X^{\mu}( \sigma+i\tau )
	=	\Xp^{\mu}( \sigma+i\tau ) + \Xm^{\mu}( \sigma+i\tau )
\, , \quad
	\tX^{\mu}( \sigma-i\tau )
	=	\tXp^{\mu}( \sigma-i\tau ) + \tXm^{\mu}( \sigma-i\tau )
\, , \\[.4em]
\label{EQN:Holomorphic_Functions}
\left\{ \;
	\begin{split}
		\Xp^{\mu}( \sigma+i\tau )
	&	:=	\frac{1}{2} \Cx^{\mu}
		-	i \sum_{n=1}^{\infty} \frac{1}{n} \, \Step_{-n}^{\mu} \, e^{-in(\sigma+i\tau)}
	\, , \\
		\Xm^{\mu}( \sigma+i\tau )
	&	:=	- 2 \Cm^{\mu} ( \sigma + i \tau )
		+	i \sum_{n=1}^{\infty} \frac{1}{n} \, \Step_{n}^{\mu} \, e^{in(\sigma+i\tau)}
	\, ,
	\end{split}
\right.
\\
\label{EQN:AntiHolomorphic_Functions}
\left\{ \;
	\begin{split}
		\tXp^{\mu}( \sigma-i\tau )
	&	:=	\frac{1}{2}  \Cx^{\mu}
		-	i \sum_{n=1}^{\infty} \frac{1}{n} \, \tStep_{-n}^{\mu} \, e^{in(\sigma-i\tau)}
	\, , \\
		\tXm^{\mu}( \sigma-i\tau )
	&	:=	2 \Cm^{\mu} ( \sigma - i \tau )
		+	i \sum_{n=1}^{\infty} \frac{1}{n} \, \tStep_{n}^{\mu} \, e^{-in(\sigma-i\tau)}
	\, ,
	\end{split}
\right.
\end{gather*}
where we divide the left-moving and right-moving modes.
Applying the general formulae (\ref{EQN:Holomorphic_vs_Holomorphic}),
(\ref{EQN:Antiholomorphic_vs_Antiholomorphic}) and (\ref{EQN:Holomorphic_vs_Antiholomorphic})
we see immediately
\begin{align*}
	\pd_{\sigma} \X^{\mu}(\sigma+i\tau) \nlNCProd \pd_{\sigma'} \X^{\nu}(\sigma'+i\tau')
&	=	\pd_{\sigma} \X^{\mu}(\sigma+i\tau) \, \pd_{\sigma'} \X^{\nu}(\sigma'+i\tau')
\, ,
\\[.4em]
	\pd_{\sigma} \tX^{\mu}(\sigma-i\tau) \nlNCProd \pd_{\sigma'} \tX^{\nu}(\sigma'-i\tau')
&	=	\pd_{\sigma} \tX^{\mu}(\sigma-i\tau) \, \pd_{\sigma'} \tX^{\nu}(\sigma'-i\tau')
\, ,
\\[.4em]
	\pd_{\sigma} \X^{\mu}(\sigma+i\tau) \nlNCProd \pd_{\sigma'} \tX^{\nu}(\sigma'-i\tau')
& \\
& \hspace{-5em}
	=	\int_{0}^{2\pi} \!\!\! \int_{0}^{2\pi}
		\frac{\d \sigma_1 \d \sigma_2}{4 \pi \NCParam} \,
		e^{i(\sigma_1-\sigma)(\sigma_2-\sigma')/2\NCParam} \,
		\pd_{\sigma_1} \X^{\mu}(\sigma_1+i\tau) \, \pd_{\sigma_2} \tX^{\nu}(\sigma_2-i\tau')
\, .
\end{align*}
\par
Despite of one's expectation the noncommutativity does not introduce
any quantum effect to the system of only left-moving or right-moving modes,
while it deforms the product of two different components.
Hence the quantization of the string coordinates must be achieved
independent from the world sheet noncommutativity.
We will adopt the standard quantization rule in which the components satisfy the commutation relations,
\begin{equation}
\label{EQN:Quantization}
	\left[ \, \Cx^{\mu}  , \Cm^{\nu}  \, \right] = i \eta^{\mu\nu}
\ , \quad
	\left[ \, \Step_{m}^{\mu}  , \Step_{n}^{\nu}  \, \right] =
	\left[ \, \tStep_{m}^{\mu} , \tStep_{n}^{\nu} \, \right] = m \delta_{m+n,0} \, \eta^{\mu\nu}
\, ,
\end{equation}
\begin{equation}
\label{EQN:Quantization_Supplement}
	\left[ \, \Step_{n}^{\mu}, \tStep_{m}^{\nu} \, \right]
	=	0
\qquad
	\forall \, n,m \neq 0
\, ,
\end{equation}
and annihilate the vacuum;
\begin{equation}
\label{EQN:Vacuum}
	\Cm^{\mu} \, \vac
	=	0
\ , \quad
	\Step_{n}^{\mu} \, \vac = \tStep_{n}^{\mu} \, \vac = 0
\quad
	\forall \, n > 0
\ .
\end{equation}
\vspace{\baselineskip}
\par
Since the \( \nlNCProd \)-product gives rise to the mixing of the fields
\( \pd_{\sigma} \X^{\mu}(\sigma+i\tau) \) and \( \pd_{\sigma} \tX^{\nu}(\sigma-i\tau) \),
they do not commute anymore.
In order to recover the commutativity of these fields in the nature of the \( \nlNCProd \)-product,
the authors in \cite{Lizzi_Vaidya_Vitale} required the conditions
\begin{equation*}
	\Step_{n}^{\mu} \, \tStep_{m}^{\nu} = e^{2 i \NCParam nm} \, \tStep_{m}^{\nu} \, \Step_{n}^{\mu}
\qquad
	\forall \, n,m \neq 0
\, 
\end{equation*}
for all non-zero modes.
\vspace{\baselineskip}
\par
On the other hand we have already shown that the quantization of the string coordinates
must be done independent from the world sheet noncommutativity.
Moreover our purpose of this note is
to see the effect of the world sheet noncommutativity to the physical objects,
rather than to discuss the conformal symmetry.
Therefore, instead of following the argument of \cite{Lizzi_Vaidya_Vitale}, 
we simply assume (\ref{EQN:Quantization_Supplement})
and see how the deformation changes the physical quantities.
\vspace{\baselineskip}
\par
The contribution of the world sheet noncommutativity appears in the vertex operator.
Since the fields \( \X^{\mu} \) and \( \tX^{\nu} \) do not commute under the \( \NCProd \)-product,
we have to fix their ordering.
We will define the ordering rule of these fields such that the field \( \X^{\mu} \)
is always in the left of the field \( \tX^{\nu} \).
According to this rule the tachyon vertex operator, for instance, is given by
\begin{align}
\label{EQN:Vertex_Operator}
	\Vertex{k}(\sigma,\tau)
&	:=	e^{i k \cdot \Xp(\sigma+i\tau)} \, e^{i k \cdot \Xm(\sigma+i\tau)}
		\NCProd
		e^{i k \cdot \tXp(\sigma-i\tau)} \, e^{i k \cdot \tXm(\sigma-i\tau)}
\nn \\
&	=	\iint_{0}^{2\pi} \frac{\d\sigma_1 \d\sigma_2}{4\pi\NCParam} \,
		e^{i( \sigma_1 - \sigma )( \sigma_2 - \sigma )/2\NCParam} \,
		e^{i k \cdot \Xp(\sigma_1+i\tau)} \, e^{i k \cdot \Xm(\sigma_1+i\tau)} \,
		e^{i k \cdot \tXp(\sigma_2-i\tau)} \, e^{i k \cdot \tXm(\sigma_2-i\tau)}
\, ,
\end{align}
where \( \, k \!\cdot\! \X := k_{\mu} \X^{\mu} = \eta_{\mu\nu} \, k^{\mu} \X^{\nu} \)
stands for the inner product.
As an application of the vertex operator,
we evaluate the tachyon propagator of a closed string:
\begin{align}
\label{EQN:Tachyon_Propagator}
&	\cvac \, \Vertex{k}(\sigma,\tau) \nlNCProd \Vertex{k'}(\sigma',\tau') \, \vac
\nn \\
& \quad
	=	\delta(k+k') \iint_{0}^{2\pi} \!
		\frac{\d\sigma_1 \d\sigma_2}{4 \pi \NCParam} \!
		\iint_{0}^{2\pi} \!
		\frac{\d\sigma'_1 \d\sigma'_2}{4 \pi \NCParam} \;
		\Delta^{k}_{(\NCParam)}(\sigma,\tau;\sigma_1,\sigma_2|\sigma'\!,\tau';\sigma'_1,\sigma'_2)
\, ,
\end{align}
where
\begin{align}
\label{EQN:Propagator}
&	\Delta^{k}_{(\NCParam)}(\sigma,\tau;\sigma_1,\sigma_2|\sigma'\!,\tau';\sigma'_1,\sigma'_2)
\nn \\[.4em]
& \quad
	:=
		e^{ i( \sigma_1  - \sigma  )( \sigma_2  - \sigma  )/4\NCParam} \,
		e^{ i( \sigma_1  - \sigma  )( \sigma'_2 - \sigma' )/4\NCParam} \,
		e^{ i( \sigma'_1 - \sigma' )( \sigma'_2 - \sigma' )/4\NCParam} \,
		e^{-i( \sigma'_1 - \sigma' )( \sigma_2  - \sigma  )/4\NCParam} \,
\nn \\
& \qquad \; \times \;
		( \, e^{-i(\sigma_1+i\tau)} \! - e^{-i(\sigma'_1+i\tau')} \, )^{-k^2}
		( \, e^{ i(\sigma_2-i\tau)} \! - e^{ i(\sigma'_2-i\tau')} \, )^{-k^2}
\, .
\end{align}
%
%
%#################################################
\section{D-brane correlators}
It is reasonable to assume that the deformation of the physical objects
by the world sheet noncommutativity is caused only through the spacetime coordinate.
From this point of view we must express all physical objects in terms of the spacetime coordinates. 
\par
We are interested in the deformation of D-brane states by the world sheet noncommutativity.
Thus we define the boundary state, which expresses a D-brane situated at the position
\( y^{j} \, \left( j = 1,\cdots,d_{\perp} \right) \) on the \( d \)-dimensional Minkowski spacetime, by 
\begin{align}
\label{EQN:Boundary_State}
&	| \rho \rangle
	:=	B_{\rho}(0) \vac
\nn \\
&	B_{\rho}(\tau)
	:=	\int \! \frac{\d^{d_{\perp}}p}{(2\pi)^{d_{\perp}}}
		\NormalOrdering
		\exp
		\left( \,
			i \rho_{\mu\nu} \! \int_{0}^{2\pi} \frac{\d \sigma}{2\pi} \,
			\X^{\mu}(\sigma+i\tau,y) \,
			\overset{\leftrightarrow}{\pd}_{\!\sigma}
			\tX^{\nu}(\sigma-i\tau,y)
		\, \right)
		\NormalOrdering
\ .
\end{align}
The \( \rho_{\mu\nu} \) is a nonsingular matrix
which specifies the boundary condition for each direction of the brane
with the value \( -1 \) or \( 1 \) corresponding to
the Neumann or Dirichlet, respectively.
The \( d_{\perp} \) is the number of directions which are the Dirichlet boundary.
In this formula we used the notation
\(
	f { \displaystyle \mathop{\pd}^{\leftrightarrow} } g
	=	\frac{1}{2} [ f ( \pd g ) - ( \pd f ) g ]
\)
and
\begin{equation*}
	\X^{\mu}(\sigma+i\tau,y)
	:=	\begin{cases}
			\X^{j}(\sigma+i\tau) - y^{j} / 2 & \quad
			j = 1, \cdots, d_{\perp} \quad \text{Dirichlet},
		\cr
			\X^{a}(\sigma+i\tau) & \quad
			a = d_{\perp}+1, \cdots, d \quad \text{Neumann},
		\cr
		\end{cases}
\end{equation*}
the same holds for \( \tX^{\mu}(\sigma-i\tau,y) \).
We also assume that the operator \( \Cm^{\mu} \) is replaced by a \( c \)-number \( p^{j} \),
which is integrated out in the Dirichlet directions, or is put zero in the Neumann directions.
If we further impose the condition
\( \bigl[ \X^{j}(\sigma) + \tX^{j}(\sigma) \bigr] |_{\sigma=2\pi} = 0 \),
we can indeed confirm that (\ref{EQN:Boundary_State}) coincides with
the well-known form of the D-brane states (see, {\it e.g.}, \cite{DiVeccia_Liccardo}):
\begin{equation*}
	| \rho \rangle
	=	\delta^{d_{\perp}}( \Cx^{j} - y^{j} ) \,
		\exp
		\biggl( \,
			\rho_{\mu\nu}
			\sum_{n=1}^{\infty} \frac{1}{n} \, \Step_{-n}^{\mu} \, \tStep_{-n}^{\nu}
		\, \biggr)
		\vac
\, .
\end{equation*}
\vspace{\baselineskip}
\par
Our expression of the boundary state (\ref{EQN:Boundary_State}) enables us
to see the contribution of the world sheet noncommutativity,
simply by replacing the product of the fields to the \( \NCProd \)-product with the ordering
as defined before,
\begin{equation}
\label{EQN:NC_Boundary_State}
	B^{\NCProd}_{\rho}(\tau)
	:=	\int \! \frac{\d^{d_{\perp}}p}{(2\pi)^{d_{\perp}}}
		\NormalOrdering
		\exp
		\left( \,
			i \rho_{\mu\nu} \! \int_{0}^{2\pi} \frac{\d\sigma}{2\pi} \,
			\X^{\mu}(\sigma+i\tau,y)
			\NCProd \overset{\leftrightarrow}{\pd}_{\!\sigma}
			\tX^{\nu}(\sigma-i\tau,y)
		\, \right)
		\NormalOrdering
\ . 
\end{equation}
Since this corresponds to the case (\ref{EQN:Holomorphic_vs_Antiholomorphic}),
we obtain a compact expression of the deformed boundary state by using
the elliptic theta function \( \RTheta{3} \),
\begin{equation}
\label{EQN:NC_Boundary_State_by_theta3}
	B^{\NCProd}_{\rho}(\tau)
	=	\int \! \frac{\d^{d_{\perp}}p}{(2\pi)^{d_{\perp}}}
		\NormalOrdering
		\exp
		\left( \,
			i \rho_{\mu\nu} \!
			\iint_{0}^{2\pi} \frac{\d\sigma \d\sigma'}{(2\pi)^2} \,
			\X^{\mu}(\sigma+i\tau,y) \,
			\overset{\leftrightarrow}{\pd}_{\!\sigma+\sigma'}
			\tX^{\nu}(\sigma'\!-i\tau,y) \,
			\RTheta{3} \!
			\left(
				\frac{\sigma - \sigma'}{2\pi},
				\frac{2\NCParam}{\pi}
			\right)
		\, \right)
		\NormalOrdering
\ ,
\end{equation}
where we used the following integral formula,
\begin{equation*}
\label{EQN:Integral_Formula_theta3}
	\RTheta{3} \!
	\left(
		\frac{\sigma_1 \! - \sigma_2}{2\pi},
		\frac{2\NCParam}{\pi}
	\right)
	=	\frac{1}{2\NCParam}
		\int_{0}^{2\pi} \! \d\sigma \;
		e^{i(\sigma_1-\sigma)(\sigma_2-\sigma)/2\NCParam}
\, . 
\end{equation*}
Although the deformed boundary state is no longer an eigenstate of the spacetime coordinate,
there exist \( \X_{(\NCParam)}^{\mu} \) and \( \tX_{(\NCParam)}^{\mu} \) which satisfy
\begin{equation}
\label{EQN:Deformed_Eigenvectors}
	\left( \, \X_{(\NCParam)}^{j} + \tX_{(\NCParam)}^{j} \, \right)_{\! \tau=0} \NCBS{\rho}
	=	y^{j}\NCBS{\rho}
\ , \quad
	\pd_{\tau} \!
	\left( \, \X_{(\NCParam)}^{a} + \tX_{(\NCParam)}^{a} \, \right)_{\! \tau=0} \NCBS{\rho}
	=	0
\ .
\end{equation}
These are defined through the redefinition of the non-zero modes of the fields such as
\begin{align*}
&	\Step_{\NCParam,\pm{n}}^{\mu}
	:=	e^{\mp i \NCParam n^2} \Step_{\pm{n}}^{\mu}
\, , \quad
	\tStep_{\NCParam,\pm{n}}^{\mu}
	:=	e^{\mp i \NCParam n^2} \tStep_{\pm{n}}^{\mu}
\quad \;
	\forall \, n > 0
\, ,
\nn \\[.3em]
&	\X_{(\NCParam)}^{\mu}(\sigma+i\tau)
	:=	\frac{1}{2} \Cx^{\mu} - 2 \Cm^{\mu}(\sigma+i\tau)
	+	i \sum_{n=1}^{\infty}
		\frac{1}{n}
		\left( \,
			\Step_{\NCParam, n}^{\mu} \, e^{ in(\sigma+i\tau)}
		-	\Step_{\NCParam,-n}^{\mu} \, e^{-in(\sigma+i\tau)}
		\, \right)
\, ,
\nn \\
&	\tX_{(\NCParam)}^{\mu}(\sigma-i\tau)
	:=	\frac{1}{2} \Cx^{\mu} + 2 \Cm^{\mu}(\sigma-i\tau)
	+	i \sum_{n=1}^{\infty}
		\frac{1}{n}
		\left( \,
			\tStep_{\NCParam, n}^{\mu} \, e^{-in(\sigma-i\tau)}
		-	\tStep_{\NCParam,-n}^{\mu} \, e^{ in(\sigma-i\tau)}
		\, \right)
\, .
\end{align*}
The modified coefficients still satisfy the same commutation relations (\ref{EQN:Quantization})
as well as (\ref{EQN:Quantization_Supplement}):
\begin{equation*}
	\left[ \, \Step_{\NCParam,m}^{\mu}  , \Step_{\NCParam,n}^{\nu}  \, \right] =
	\left[ \, \tStep_{\NCParam,m}^{\mu} , \tStep_{\NCParam,n}^{\nu} \, \right] =
	m \delta_{m+n,0} \, \eta^{\mu\nu},\qquad
	\left[ \, \Step_{\NCParam,m}^{\mu}, \tStep_{\NCParam,n}^{\nu} \, \right]
	=	0
\, .
\end{equation*}
\vspace{\baselineskip}
\par
Now we want to know the effect of the world sheet noncommutativity in physical objects.
For example let us study a correlation function with respect to
the deformed D-branes (\ref{EQN:NC_Boundary_State}).
The tachyon scattering off a D-brane is evaluated by the following expectation value
\begin{equation*}
\label{EQN:Expectation_Value}
	\cNCBS{\rho,\tau}
		\, \Vertex{k}(\sigma,\tau) \nlNCProd \Vertex{k'}(\sigma',0) \,
	\NCBS{\rho',0}
	=	\cvac \, B^{\ast}_{\rho}(\tau)
		\, \Vertex{k}(\sigma,\tau) \nlNCProd \Vertex{k'}(\sigma',0) \,
		B^{\ast}_{\rho'}(0) \, \vac
\ .
\end{equation*}
The vacuum expectation value on the right hand side can be evaluated easily
if we rewrite the boundary state (\ref{EQN:NC_Boundary_State}) by using auxiliary fields as
\begin{align}
\label{EQN:Linearized_Boundary_State}
	B^{\ast}_{\rho}(\tau)
&	=	B_{0}(y^{j}) \iint \! \mathcal{D} \Aux \, \mathcal{D} \tAux \,
		\exp
		\Bigl[ \,
			-i \!
			\iint_{0}^{2\pi}
			\frac{\d \sigma \d \sigma'}{(2\pi)^2} \,
			\rho^{\mu\nu}
			\tAux_{\mu}(\sigma) \, \pd_{\sigma'} \Aux_{\nu}(\sigma') \,
			\RTheta{3} \! \left( \frac{\sigma-\sigma'}{2\pi}, - \frac{2 \NCParam}{\pi} \right)
		\, \Bigr]
\nn \\
& \quad \times \;
		\NormalOrdering
		\exp
		\Bigl[ \,
		-	i \! \int_{0}^{2\pi} \! \frac{\d \sigma}{2\pi} \,
			\Aux_{\mu}(\sigma) \, \pd_{\sigma} \X^{\mu}(\sigma+i\tau)
		+	i \! \int_{0}^{2\pi} \! \frac{\d \sigma}{2\pi} \,
			\tAux_{\mu}(\sigma) \, \pd_{\sigma} \tX^{\mu}(\sigma-i\tau)
		\, \Bigr]
		\NormalOrdering
\, ,
\end{align}
where \( \rho^{\mu\nu} \) is the inverse matrix of \( \rho_{\mu\nu} \, \)
and \( B_{0}(y^{j}) \) is an operator which specifies the position of the D-brane
which is not affected by the world sheet noncommutativity.
Note that 
\(
	\int_{0}^{2\pi} \d \sigma \, \Aux_{\mu}(\sigma)
	=	\int_{0}^{2\pi} \d \sigma \, \tAux_{\mu}(\sigma)
	=	0
\)
holds because the zero modes of the string coordinate vanish on the vacuum. 
\par
To proceed further we introduce a modified theta function
\begin{equation*}
\label{EQN:Modified_theta3}
	\Theta( \sigma, \tau, \NCParam )
	:=	\sum_{n = -\infty}^{\infty}
		e^{in\sigma} e^{-|n|\tau} e^{- 2 i \NCParam n^2}
\, , \quad
	\Theta( \sigma, 0, \NCParam )
	=	\RTheta{3} \! \left( \frac{\sigma}{2\pi}, -\frac{2\NCParam}{\pi} \right)
\, ,
\end{equation*}
and define a function \( G \) with \( \varepsilon, \beta \in \R \) by
\begin{equation}
\label{EQN:Inverse_of_theta3}
	G^{(\varepsilon)}_{\beta}( \sigma, \tau, \NCParam )
	:=	- i \sum_{n=1}^{\infty} \frac{1}{n} \! \cdot \!
		\frac{e^{in\sigma} + \varepsilon e^{-in\sigma}}%
		     {1 - \beta \, e^{- 2 i \NCParam n^2} e^{n \tau}}
	+	\frac{c(\varepsilon,\beta)}{1 - \beta} \, \sigma
\; , \quad
	c(\varepsilon,\beta)
	=	\left\{
		\begin{array}{cl}
			1 & \varepsilon = -1, \beta \neq 1 \cr
			0 & \text{otherwise}
		\end{array}
		\right.
\end{equation}
such that it obeys
\begin{equation}
\label{EQN:Integral_Formula_for_G_Inverse}
	\int_{0}^{2\pi} \! \frac{\d \sigma''}{2\pi}
	\left[ \,
		2 \pi \delta(\sigma-\sigma'')
	-	\beta \, \Theta( \sigma-\sigma''\!, -\tau, \NCParam )
	\, \right]
	\pd_{\sigma''} G_{\beta}^{(-1)}( \sigma''\!\!-\sigma', \tau, \NCParam )
	=	2 \pi \delta( \sigma-\sigma' )
\, ,
\end{equation}
in the case \( \varepsilon = -1 \).
These are sufficient to calculate the Gaussian integration over the auxiliary fields.
\par
For the sake of simplicity, let us specify the boundary conditions to the case
\(
	( \rho_{\mu\nu} ) = ( \rho'_{\mu\nu} )
	=	\mathrm{diag}( \delta_{jk}, - \eta_{ab} )
\).
Then the expectation value results in 
\begin{align}
\label{EQN:Evaluated_Expectation_Value}
&	\frac{
		\cNCBS{\rho,\tau}
			\, \Vertex{k}(\sigma,\tau) \nlNCProd \Vertex{k'}(\sigma',0) \,
		\NCBS{\rho',0}
	}{
		\NCBracket{\rho,\tau}{\rho',0}
	}
\nn \\[.2em]
&	=	\delta(k+k') \iint_{0}^{2\pi} \!
		\frac{\d\sigma_1 \d\sigma_2}{4 \pi \NCParam} \!
		\iint_{0}^{2\pi} \!
		\frac{\d\sigma'_1 \d\sigma'_2}{4 \pi \NCParam} \;
		\Delta^{k}_{(\NCParam)}(\sigma,\tau;\sigma_1,\sigma_2|\sigma'\!,\tau';\sigma'_1,\sigma'_2)
\nn \\
& \quad \times \,
		\biggl(
			\frac{ \Phi_{(\NCParam)}^{+}(0,\tau)^2 }%
			     { \Phi_{(\NCParam)}^{+}(\sigma_1-\sigma'_1+i\tau,\tau)
			       \Phi_{(\NCParam)}^{+}(\sigma_2-\sigma'_2-i\tau,\tau) }
		\biggr)^{\! - k^2}
\nn \\
& \quad \times \,
		\biggl(
			\frac{ \Phi_{(\NCParam)}^{-}(\sigma'_1-\sigma_2,\tau)
			       \Phi_{(\NCParam)}^{-}(\sigma_1-\sigma'_2,\tau) }%
			     { \Phi_{(\NCParam)}^{-}(\sigma_1-\sigma_2+i\tau,\tau)
			       \Phi_{(\NCParam)}^{-}(\sigma'_1-\sigma'_2-i\tau,\tau) }
		\biggr)^{\! - k \cdot S \cdot k}
\, ,
\end{align}
where we used the definition (\ref{EQN:Propagator}) and denoted that
\begin{equation*}
	\Phi_{(\NCParam)}^{\pm}(\sigma,\tau)
	:=	\exp
		\left( \,
			\frac{i}{2}
			\left[ \,
				G^{(1)}_{1}(\sigma,\tau,\NCParam) \pm G^{(1)}_{-1}(\sigma,\tau,\NCParam)
			\, \right]
		\, \right)
\, , \qquad
	S
	=	\left(
			\begin{array}{cc}
				\mathbf{1}_{d_{\perp}} & 0 \cr
				0 & - \mathbf{1}_{d_{\parallel}}
			\end{array}
		\right)
\, ,
\end{equation*}
and \( d_{\parallel} = d - d_{\perp} \) is the number of dimensions dominated by the D-brane.
\vspace{\baselineskip}
\par
From the analysis we carried out in this section
we can easily read off some of the effects of the world sheet noncommutativity to the D-brane correlators. 
For example, the change of the free energy part of the D-branes,
which can be obtained by the calculation of
\( \cvac \, B^{\ast}_{\rho}(\tau) \, B^{\ast}_{\rho'}(0) \, \vac \),
is simply given by the following modification of the Dedekind eta function:
\begin{equation}
\label{EQN:Eta_Function}
	\eta(q)
	=	q^{\frac{1}{24}}
		\prod_{n=1}^{\infty}
		\left(
			1 - q^{n}
		\right)
	\quad \xrightarrow[\NCParam]{\hspace{3em}} \quad
		q^{\frac{1}{24}}
		\prod_{n=1}^{\infty}
		\left(
			1 - e^{4 i \NCParam n^2} q^{n}
		\right)
\quad \text{where} \quad
	q = e^{-2\tau}
\, .
\end{equation}
\par
The effect of the world sheet noncommutativity to the tachyon propagator
between the D-branes are separated into two parts. 
One is the free tachyon propagator \( \Delta^{k}_{(\NCParam)} \).
The deformation of this part is caused by the appearance of the \( \nlNCProd \)-product
in all vertex operators and their products, and is nothing to do with the boundary states. 
\par
The other part of the effects of the world sheet noncommutativity is
in \( \Phi^{\pm}_{(\NCParam)} \) of (\ref{EQN:Evaluated_Expectation_Value}).
In contrast to the free tachyon propagators we notice that
their \( \NCParam \)-dependence comes only from the deformation of the D-brane states
\( B^{\ast}_{\rho}(\tau) \) of (\ref{EQN:NC_Boundary_State_by_theta3}).
In the derivation of \( \Phi^{\pm}_{(\NCParam)} \)
the \( \NCProd \)-products between tachyon vertices are irrelevant. 
Therefore we are able to deal with the effects of the world sheet noncommutativity
on the vertex operators and the D-brane states, independently.
\par
In fact we can derive the commutative limit
\( \NCParam \to 0 \) of (\ref{EQN:Evaluated_Expectation_Value}) as
\begin{align}
\label{EQN:Commutative_Limit}
	\lim_{\lambda \, \to \, 0}
&	\frac{
		\cNCBS{\rho,\tau}
			\, \Vertex{k}(\sigma,\tau) \nlNCProd \Vertex{k'}(\sigma',0) \,
		\NCBS{\rho',0}
	}{
		\NCBracket{\rho,\tau}{\rho',0}
	}
\nn \\
&	=	\delta(k+k') \,
		e^{- 2 k \cdot S \cdot k \, \zeta(1)} \,
		e^{- 2 \tau ( k^2 + k_{\parallel}^2 )}
		\left| \,
			\frac{
				\RTheta{1}^{\prime}( 0, \frac{i\tau}{\pi} ) / 2 \pi
			}{
				\RTheta{1}( \frac{\sigma+i\tau-\sigma'}{2\pi}, \frac{i\tau}{\pi} )
			}
		\, \right|^{- 4 k_{\parallel}^2}
		\left| \, 1 - e^{i(\sigma+i\tau-i\sigma')} \, \right|^{- 4 k^2}
\, ,
\end{align}
where \( 2 k_{\parallel}^2 := k^2 - k \!\cdot\! S \!\cdot\! k \).
This agrees with the result which we obtain by the calculation of the same object
by putting \( \NCParam = 0 \) in the \( \Delta^{k}_{(\NCParam)} \) of (\ref{EQN:Propagator})
and in the \( \RTheta{3} \) function of (\ref{EQN:NC_Boundary_State_by_theta3}).
%
%
%#################################################
\section{Conclusion}
By introducing the noncommutativity in the world sheet,
we discussed a modification of the D-brane states in the closed string theory.
In particular we have shown, in this note,
how the world sheet noncommutativity induces a nonperturbative effect to the D-brane states.
The results were based on our assumption that the conventional quantization rules
(\ref{EQN:Quantization}) and (\ref{EQN:Quantization_Supplement}) are not changed
by the world sheet noncommutativity.
The world sheet noncommutativity induces the noncommutativity of only the product of the fields
\( \pd_{\sigma} \X^{\mu}(\sigma+i\tau) \) and \( \pd_{\sigma} \tX^{\mu}(\sigma-i\tau) \).
\par
This assumption owes to the fact that the \( \NCProd \)-product (\ref{EQN:Nonlocal_Star_Product})
does not give rise to the quantization of the spacetime.
We must quantize the string coordinate independently from the world sheet noncommutativity.
Once we adopt this assumption as our starting point of our discussion
we can evaluate the effects of the world sheet noncommutativity nonperturbatively.
For example we found that the D-brane boundary states are modified simply
replacing the \( \delta \) function by the elliptic function \( \RTheta{3} \)
whose modulus is the noncommutativity parameter \( \NCParam \) itself.
This phenomenon can be interpreted
as a consequence of the topology change of the world sheet from a sphere to a torus.
We emphasize that it will be difficult to see this result by the perturbative calculations.
\par
Meanwhile we have found that the deformed D-brane states have
a nice feature given by (\ref{EQN:Deformed_Eigenvectors}) in contrast to one's misgivings.
Namely the modification of the D-branes is equivalent to the deformation of the string coordinates
\( \X^{\mu} \to \X_{(\NCParam)}^{\mu} \).
Moreover the simple formula of the modified D-brane states enables us to evaluate
D-brane correlators nonperturbatively, as we have shown in section 3.
The result (\ref{EQN:Evaluated_Expectation_Value}) tells us how
the world sheet noncommutativity appears in the D-brane physics.
In particular the \( \NCParam \)-dependence of the D-brane states appears in
the correlators only through the function \( G^{(\varepsilon)}_{\beta}(\sigma,\tau,\NCParam) \)
of (\ref{EQN:Inverse_of_theta3}).
\vspace{\baselineskip}
\par
We have not discussed about the breaking of the conformal symmetry
by the introduction of the noncommutativity parameter \( \NCParam \).
In order to clarify this problem within the frame work of the quantum field theory, however,
we must specify a model to be considered and
study the relationship between the symmetry and the quantization procedure.
Our results are expected to contribute to a better understanding of such problems.
\par
Since the purpose of this note is to show
the possibility of evaluating the nonperturbative effects of the world sheet noncommutativity,
we have ignored not only the supersymmetric partners but also the ghosts.
In order to discuss physical insights of our results we must take them all into account,
which we are going to discuss in our forthcoming paper.
%
%
%#################################################
\section*{Acknowledgements}
The author would like to thank Dr.~Satoru Saito, Dr.~Tomoya Hatanaka and Dr.~Noriaki Kitazawa
for their encouragement and helpful discussions.
%
%
%#################################################

%
%
\end{document}